\documentclass[lettersize,journal]{IEEEtran}
\IEEEoverridecommandlockouts
\usepackage{cite}
\usepackage{amsmath,amssymb,amsfonts}
\usepackage{bm}
\usepackage{setspace}
\usepackage{graphicx}
\usepackage{CJKutf8}
\usepackage{textcomp}
\usepackage{xcolor}
\usepackage{algorithm} 
\usepackage{algpseudocode} 
\usepackage{comment}
\usepackage{subcaption}
\newcommand*{\note}[1]{\textcolor{red}{#1}}
\usepackage{mathrsfs}
\usepackage{xspace}
\usepackage{graphicx}
\usepackage{multirow}
\usepackage{threeparttable}
\usepackage[normalem]{ulem}
\newcommand*{\shen}[1]{{\textcolor{black}{#1}}}
\newcommand{\sysname}{$\text{DDRO}$\xspace}

\def\BibTeX{{\rm B\kern-.05em{\sc i\kern-.025em b}\kern-.08em
    T\kern-.1667em\lower.7ex\hbox{E}\kern-.125emX}}
\begin{document}


\title{Intelligent Task Management via Dynamic Multi-region Division in LEO Satellite Networks}



\author{Zixuan Song, Zhishu Shen,~\IEEEmembership{Member, IEEE}, Xiaoyu Zheng, Qiushi Zheng,~\IEEEmembership{Member, IEEE}, Zheng Lei, and Jiong~Jin,~\IEEEmembership{Member,~IEEE}
\thanks{Zixuan Song, Zhishu Shen, Xiaoyu Zheng are with School of Computer Science and Artificial Intelligence, Wuhan University of Technology, Wuhan, China (e-mail: putangan123@whut.edu.cn, z\_shen@ieee.org, 311431@whut.edu.cn). Zhishu Shen is also with the Hubei Key Laboratory of Transportation Internet of Things, Wuhan University of Technology, Wuhan, China.}
\thanks{Qiushi Zheng, Zheng Lei, and Jiong Jin are with the School of Engineering, Swinburne University of Technology, Melbourne, Australia (e-mail: \{qiushizheng, zlei, jiongjin\}@swin.edu.au)}
\thanks{This work was supported in part by the National Natural Science Foundation of China (Grant No. 62472332) and the Hubei Provincial International Science and Technology Cooperation Project (No. 2024EHA031).} 
\thanks{\textit{Corresponding author: Zhishu Shen.}}}



\maketitle

\begin{abstract}
As a key complement to terrestrial networks and a fundamental component of future 6G systems, Low Earth Orbit (LEO) satellite networks are expected to provide high-quality communication services when integrated with ground-based infrastructure, thereby attracting significant research interest. However, the limited satellite onboard resources and the uneven distribution of computational workloads often result in congestion along inter-satellite links (ISLs) that degrades task processing efficiency. Effectively managing the dynamic and large-scale topology of LEO networks to ensure balanced task distribution remains a critical challenge. To this end, we propose a dynamic multi-region division framework for intelligent task management in LEO satellite networks. This framework optimizes both intra- and inter-region routing to minimize task delay while balancing the utilization of computational and communication resources. Based on this framework, we propose a dynamic multi-region division algorithm based on the Genetic Algorithm (GA), which adaptively adjusts the size of each region based on the workload status of individual satellites. Additionally, we incorporate an adaptive routing algorithm and a task splitting and offloading scheme based on Multi-Agent Deep Deterministic Policy Gradient (MA-DDPG) to effectively accommodate the arriving tasks. Simulation results demonstrate that our proposed framework outperforms comparative methods in terms of the task delay, energy consumption per task, and task completion rate.

\end{abstract}


{
\section{Introduction}

With the development of emerging Internet of Things (IoT) and Artificial Intelligence (AI) technologies, computationally intensive tasks such as intelligent driving and video analysis are placing increasing demands on communication systems. However, in remote rural areas, the lack of communication infrastructure, such as base stations, often prevents local data transmission and processing from being fully  executed~\cite{fang20215g,azari2022evolution}. In the era of big data, efficiently handling computationally intensive tasks originating from these underserved regions has emerged as a critical challenge~\cite{yang2023fhap}. Compared to terrestrial communication, satellite communication offers unique advantages by providing global coverage and extending connections to areas beyond the reach of traditional terrestrial networks~\cite{shen2023survey,LiuTNSM24}. Especially, the Low Earth Orbit (LEO) satellite networks, composed of massive LEO satellites, are increasingly being deployed to handle a wide range of tasks~\cite{HuTNSM22,ZhangTNSE23,LiuTNSM242}.

With the continuous increase in the number of LEO satellites, the network structure is becoming increasingly complex and dynamic. The high-speed movement of satellites causes the LEO satellite network topology to constantly change, which poses a huge challenge to establishing stable and reliable point-to-point communication links between LEO satellites~\cite{jiang2021qoe}. 
To maintain the continuity and data integrity of communication connections, current research tends to implement topology region division strategies for networks to reduce the complexity of routing calculations~\cite{zhou2024intelligent,FangTWC23}. This strategy is founded on the distribution of communication demand corresponding to the ground population density or the topological characteristics exhibited by satellite networks.  By clustering satellites in adjacent orbits and defining spatial areas, it ensures stable inter-regional routing and minimizes disruptions caused by frequent satellite link changes~\cite{zhang2021aser}. 

Based on the regional division of LEO satellite networks, the routing mechanism is composed of two categories: static routing and dynamic routing~\cite{han2022time}. The static routing relies on virtual topology and virtual nodes to address the dynamic nature of satellite network topology. It divides time into discrete slices representing static scenarios and computes routes sequentially within each slice~\cite{chen2024shortest}. Although static routing is simple to implement, it often overlooks network load balancing, which can result in congestion. In contrast, the dynamic routing adapts to real-time topology changes, offering a solution to alleviate network congestion~\cite{wang2021fuzzy}. However, the current solution necessitates extensive exchange of link state information within the network, potentially resulting in network congestion and high resource consumption~\cite{liu2024multipath}. To address these challenges, recent research has focused on designing routing algorithms that strike a balance between computational complexity and routing optimization, rather than finding strictly optimal routes. This strategy aims to minimize redundant overhead during message transmission, prevent real-time congestion, and enhance load balancing efficiency~\cite{JiaoTVT24,WangIoT25}. However, this approach overlooks the influence of network dynamics on regional partitioning. As satellites in LEO networks move continuously along their orbits, some satellites initially assigned to a specific region may drift over time~\cite{liu2021reliable}. Outdated regional divisions can increase communication overhead and reduce the effectiveness of routing algorithms.

To this end, we propose a framework for Dynamic multi-region Division, Routing and Offloading, named \sysname, for LEO satellite networks. This framework initially divides the network into multiple equal-sized regions and aims to process incoming tasks within each region, thereby maximizing the utilization of satellite resources locally. By analyzing the load conditions within each sub-region, our framework can dynamically adjust regional boundaries to alleviate congestion in overloaded areas. To support this dynamic adjustment, we design a neighborhood information exchange strategy that facilitates efficient coordination between adjacent regions. Based on this dynamic multi-region division framework, we further propose a routing and task offloading scheme that optimizes both intra- and inter-region routing, aiming to reduce routing complexity and maximizing the utilization of communication and computation resources. In this way, our proposal effectively mitigates network congestion in high traffic load areas under dynamic network conditions.  The primary contributions of our work include: 


\begin{enumerate}

\item We introduce a dynamic multi-region division framework for intelligent task management in resource-constrained LEO satellite networks. 
The multi-region division process involves a \shen{region division} and region re-division using a Genetic Algorithm (GA)-based algorithm. This adaptive process is supported by a neighborhood information exchange strategy, which enables flexible multi-region adjustment to maintain load balance under dynamic network conditions.

\item Based on this framework, we design a scheme for dynamic multi-region division, routing and task offloading. The adaptive routing algorithm tries to select the optimal route within a specific region in real time to accommodate network dynamics. Meanwhile, a task splitting and offloading algorithm based on Multi-Agent Deep Deterministic Policy Gradient (MA-DDPG) divides each incoming task into multiple segments. These segments are then processed collaboratively by satellites along the selected route, aiming to reduce the average task delay and task failure rate.






\item We develop a satellite network testbed to validate the proposed dynamic multi-region division, adaptive routing, and task offloading scheme. Comprehensive experiments conducted across varying network scales demonstrate that our proposal outperforms baseline methods in terms of task completion rate, task delay, energy consumption, and resource utilization.


\end{enumerate}

The remainder of this paper is organized as follows: Section II summarizes the related work, and Section III describes the problem statement. Section IV presents our proposed dynamic multi-region division, routing and task offloading scheme. Section V summarizes the evaluation results that verify the performance of \shen{the proposed framework}. Section VI gives the conclusions with our future work.

\section{Related Work}

\subsection{Region Division in Satellite Networks}

Currently, there are two main methods for dividing LEO satellite network regions. One is based on uneven division of ground conditions, such as population density and communication demand. The most fundamental way is to evenly divide the Earth's surface through latitude and longitude lines, thereby achieving the dividing of LEO satellite networks~\cite{feng2024distributed}. Specifically, Jiao \textit{et al}. proposed an adaptive satellite cluster partitioning method based on moving direction, line-of-sight link, and latitude. By dynamically adjusting the cluster size and structure, the large-scale LEO constellation is partitioned into stable subregions to reduce routing computation overhead and optimize multi-mode data transmission efficiency~\cite{jiao2024clustered}. Mao \textit{et al}. divided ground users based on their latitude and longitude and established a mapping relationship between LEO satellite regions and ground user regions, thereby completing the dividing of the LEO network and reducing the complexity of managing large-scale LEO satellite networks~\cite{mao2024intelligent}.

Another method is rectangular division based on the LEO network topology structure: Li \textit{et al}. divided the network into multiple regions based on its topological characteristics. The regions are not fixed, but at least two ISLs are included to ensure insensitivity to single link faults; Each region is further divided into four quadrants, forming horizontal and vertical transmission clusters~\cite{li2023leo}. Similarly, Zhang \textit{et al}. used rectangular uniform dividing to ensure the stability of communication links. Based on this division, the route is reconstructed to minimize the overhead~\cite{zhang2021aser}.

By implementing region division mechanism, the large-scale routing problem is mitigated by creating the routes within a small specific area. This allows for distributed routing between regions to be managed through edge nodes, thereby reducing the computational overhead. However, the dynamic impact of satellite movement on the network structure within the region remains unresolved. Due to the continuous motion of satellites, the network structure undergoes constant changes, and such frequent variations significantly affect communication efficiency.



\subsection{Task Offloading in Satellite Networks}
In the domain of LEO satellite networks, researchers have proposed various approaches to tackle the issues of task offloading: 
Zhang \textit{et al}. introduced a satellite task offloading method based on multi-agent Deep Reinforcement Learning (DRL), which uses counterfactual multi-agent strategy gradient and improved bidirectional short-term memory network to achieve energy efficient task offloading and resource allocation~\cite{zhang2024collaborative}. Lai \textit{et al}. proposed a computing offloading scheme for LEO satellite networks based on hierarchical multi-agent DRL. The experiments show the scheme's effectiveness in task processing delay, overrun rate and load balancing~\cite{lai2024joint}. These studies assume the computational load of user tasks to be uniform, permitting offloading based on percentages. However, in reality, various user tasks comprise several interdependent sub-tasks, with each sub-task being the smallest logical execution unit~\cite{wang2021dependent,PengTNSE25}. 

To overcome the aforementioned issues, Cao \textit{et al}. tackled the dependent task offloading problem in multi-user, multi-party scenarios. They employed Directed Acyclic Graphs (DAGs) to represent dependent tasks, where nodes signify sub-tasks and directed edges indicate the dependencies among them. They introduced a scheme leveraging Graph Attention Networks (GAT) and DRL to minimize time overhead~\cite{cao2024dependent}. Similarly, Chen \textit{et al}. proposed a DRL-based algorithm, which can jointly optimize service migration and power control to solve the task offloading challenge in LEO satellite networks. The experimental results validate its efficiency in terms of task completion rate and energy efficiency~\cite{chen2024spaceedge}. Chai \textit{et al}. proposed a low-complexity multi-task dynamic offloading scheme. By modeling tasks as DAGs and converting them into one-dimensional vectors, a training method combining attention mechanisms with the Proximal Policy Optimization (A-PPO) algorithm was utilized to achieve joint offloading of multi-task systems cost-efficiently~\cite{chai2023joint}. Zhang
et al. proposed a developed DDPG-based task offloading scheme to minimize the system energy consumption for satellite networks~\cite{zhang2023satellite}. However, the aforementioned algorithms adopt a global perspective and fail to account for the viewpoint of individual satellites, especially in the perspective of load balancing among the satellites. 





\section{Problem Statement}



\subsection{System Model}
We consider a satellite-terrestrial network composed of LEO satellites and ground base stations. In this network, $N$ satellites are evenly distributed on $P$ circular and evenly spaced polar orbital planes. The satellites and planes are denoted as $N$ = $\{1,2,\cdots,N\}$, and $P$ = $\{1,2,\cdots,P\}$, respectively. Each orbital plane contains $N_{p}$ satellites, where $N_{p}$=$N$/$P$. Each satellite has four ISLs, consisting of two inter-orbital links and two intra-orbital links to offload the arriving task to its neighbors. \shen{Here, the connections between satellites within the same orbital plane are referred to as intra-orbital links, whereas those between satellites in adjacent orbital planes are  inter-orbital links.} Ground stations dispatch a sequence of pending tasks to the nearest LEO satellite. Let $v_{a}$ = $\{v_{1},v_{2},\cdots,v_{N}\}$, where $v_{a}$ represents a satellite and $a \in N$. 

\shen{In this paper, source satellites refer to LEO satellites that directly receive tasks transmitted from ground stations. These tasks support splitting, allowing distributed execution across multiple collaborating satellites to enhance processing efficiency.} Let $q_{m}$ denote the size of the $m$-th task and let $q_{m}$ = $\{q_{m,1},\cdots,q_{m,k},\cdots,q_{m,K}\}$, where $q_{m,k}$ represents the size of the $k$-th sub-task of the $m$-th task. Due to the symmetric and regular grid structure of LEO satellite networks, we adopt a region division strategy that divides the satellite network topology into multiple regions. This division is intended to enhance the capability to process tasks locally within specific regions, \shen{thereby improving overall task handling efficiency.}

\subsection{Computation and Communication Model}
The total delay for offloading task $m$ to a set of satellites $\{v_{1},v_{2},\cdots,v_{n}\}$ includes computation delay and communication delay. The total delay of task $m$ can be defined as:
\begin{equation}
   \mathit{t_{m} = t_{m}^{comp} + t _{m}^{comm}}  
\end{equation}

The computation delay associated with task execution depends on the size of the task data and the available computing resources of the satellite at the time. We assume that the first satellite receiving the task is responsible for splitting it into multiple sub-tasks and offloading them to different LEO satellites for further processing. For each divided \shen{sub-task}, let $q_{m,k}$ denote the \textcolor{black}{data} size of the $k$-th sub-task of task $m$ and let $K$ represent the total number of sub-tasks. The computation delay $t_{m, k}^{comp}$  for the $k$-th sub-task of task $m$ is calculated as follows:

\begin{equation}
    t_{m,k}^{comp} = \frac{ \eta \cdot q_{m,k}}{f_{m,k}}
\end{equation}
where $\eta$ is the processing density, defined as the number of CPU cycles required to process one byte of data and $f_{m,k}$ represents the computation resources allocated to the $k$-th sub-task of the task $m$.

\begin{figure*}[tb!]
    \centering
    \includegraphics[width=0.85\textwidth]{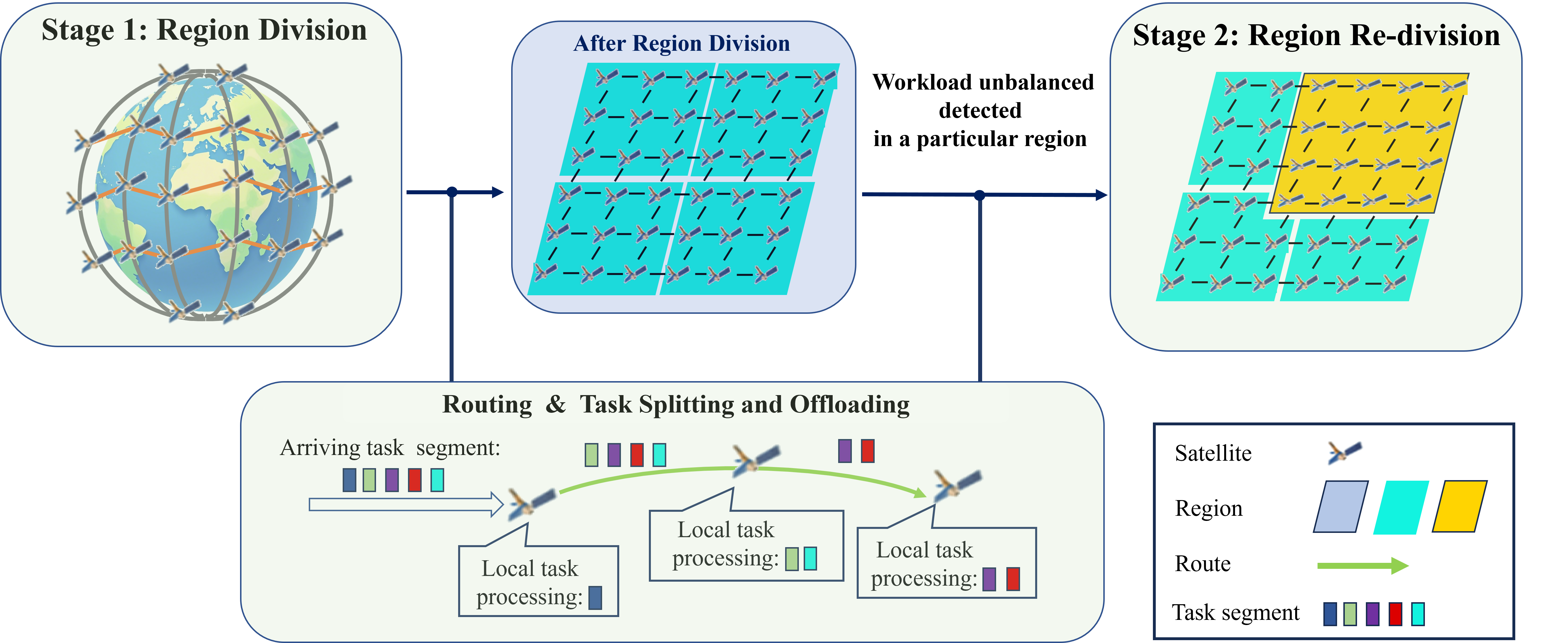}
    \caption{Overview of the proposed framework \sysname. }
    \label{fig:framework}
\end{figure*}

The communication delay is determined by both transmission delay and propagation delay. Before completing task segmentation and offloading decision on the satellite, the base station $b$ needs to collect tasks from its area and establish a stable communication link with a LEO satellite $i$. The transmission rate for this link can be calculated as follows:
\begin{equation}
   v_{b,i}(t)=B_0\log_{2}{(1+\frac{P_b\xi_{b,i}(t)}{M_G})}
\end{equation}
where $B_0$ is the channel bandwidth, $P_b$ represents the transmit power, $\xi_{b, i}(t)$ denotes the channel gain, which consists of large-scale fading and shadowed-Rician fading\cite{tang2021computation}, and $M_G$ indicates the power of additive white Gaussian noise.

Due to communication distance limitations, each satellite can forward tasks only to its neighboring satellites via inter-satellite links (ISLs). Assuming Gaussian channels, the maximum achievable data rate~\cite{mayorga2021inter} for the transmission between satellites $i$ and $j$ is given by:
\begin{equation}
    r(i,j)=B\log_{2}{(1+\frac{P_tG_i(j)G_j(i)L_i(j)L_j(i)}{kTB})}
\end{equation}
where $B$ is the bandwidth between satellites, $P_t$ is the transmission power, and $G_i(j)$ and $G_j(i)$ denote the gains of the transmitting and receiving antennas, respectively. $L_i(j)$ and $L_j(i)$ represent the beam pointing loss coefficient for the transmitter and receiver ($L_i(j), L_j(i) < 1$), $k$ is the \shen{Boltzmann constant}, and $T$ is the \shen{equivalent noise temperature.}

The transmission delay is influenced by task size and the allocated 
communication resources. Assuming that the transmission route for the $k$-th sub-task of task $m$ is $R=\{(v_1, v_2),(v_2, v_3), ..., (v_{a-1}, v_{a})\}$, the communication delay $t_{m,k}^{trans}$ for this sub-task can be calculated as follows:
\begin{equation}
    t_{m,k}^{comm} = \frac{q_{m,k}}{r_{m,k}} + \sum_{\forall  (v_i,v_j)\in R} \frac{d_{i,j}}{v_{c}}
\end{equation}
where $r_{m,k}$ represents the communication resources allocated to the sub-task and $d_{i,j}$ represent the distance between satellite $v_i$ and satellite $v_j$.  $v_c$ is the light speed.

A task is considered complete only when all of its sub-tasks have finished execution. The total delay of \shen{tasks} is calculated as follows:
\begin{equation}
    t_{m} = \max_{1 \leq k \leq K}(t_{m,k}^{comp}+t_{m,k}^{comm})
\end{equation}



\subsection{Energy consumption Model and Completion rate}

The energy consumption consists of computation consumption and communication consumption:
\begin{equation}
    e =e_{}^{comp}+e_{}^{comm} 
\end{equation}

The computation energy consumption for the $k$-th sub-task of task $m$ is given by: 
\begin{equation}
    e_{m,k}^{comp} = \kappa_0 \cdot \eta \cdot q_{m,k} \cdot f_{m,k}^{2} 
\end{equation}
where $\kappa_0$ is a chip-dependent coefficient.

The communication energy consumption for the $k$-th sub-task of task $m$ is given by: 
\begin{equation}
    e_{m,k}^{comm} = \sum_{\forall  (v_i,v_j)\in R} P_t \cdot \frac{q_{m,k}}{r_{i,j}}
\end{equation}

The total energy consumption for task $m$ is calculated as follows:
\begin{equation}
    e_{m} = \sum_{k=1}^{K} (e_{m,k}^{comp}+e_{m,k}^{comm})
\end{equation}

\shen{If a sub-task assigned to a satellite is dropped due to a lack of computational or communication resources, all subsequent sub-tasks will be terminated. }

Let $D_{i,m}$ indicates the execution result of the $m$-th task on source satellite 
$v_i$. If the task is completed, then 
$D_{i,m}$=1; otherwise, 
$D_{i,m}$=0., the total completion rate $b_{m}$ can be calculated as:
\begin{equation}
    b_{m} = \frac{\sum_{i=1}^{N_{src}}\sum_{m=1}^{M}D_{i,m}}{N_{src} \cdot M},D_{i,m} \in \{0,1\}
\end{equation}
where $N_{src}$ represents the number of source satellites, and $M$ is the total number of tasks on each source satellite.

\subsection{Optimization Objective}~\label{sec:optimizer}

As mentioned above, the offloading cost primarily consists of task delay and drop rate, which depend on offloading locations, computational resources and bandwidth allocation. Therefore, the offloading cost minimization problem $\mathcal{P}$ is formulated as follows:
\begin{equation}
    min(\overline{t}-\omega {b_m})
    \label{eq:obj}
\end{equation}
s.t.
\begin{equation}
    \sum f_{i,m,n}(t)<f_{max}^{i}
\end{equation}
\begin{equation}
    \sum r_{i,m,n}(t)<r_{max}^{i}
\end{equation}
\begin{equation}
    \sum e_{i,m,n}(t) <E_{i}(t)
\end{equation}
\begin{equation}
    \sum x _{m,n} =1
\end{equation}
where $\overline{t}$ denotes the average task delay, and $\omega$ is a weight coefficient. The terms $f_{i,m,n}(t)$ , $r_{i,m,n}(t)$ and $e_{i,m,n}(t)$ denote the computational resource consumption, communication resource consumption and stored energy for the $n$-th segment of the $m$-th task on the $i$-th satellite at time $t$. $f_{max}^{i}$ and $r_{max}^{i}$ denote the maximum available computational and communication resources of the satellite $i$. $E_{i}(t)$ denotes the remaining energy of satellite $i$ at time $t$. $x _{m}^{n}$ denotes the ratio of the $n$-th sub-task of the task $m$.


This problem is commonly reformulated using specific relaxation approaches and solved via powerful convex optimization methods. However, these approaches typically require extensive iterations, making them unsuitable in real-world systems. To this end, we are motivated to design an efficient multi-region division, routing, and task offloading scheme with low computational complexity to obtain a near-optimal solution.


\section{Dynamic Multi-region Division Framework}

\subsection{Overview}

To address the optimization problem described in Section~\ref{sec:optimizer}, we propose a dynamic multi-region division framework \sysname for LEO satellite networks. 
\figurename~\ref{fig:framework} illustrates an overview of our framework, which includes dynamic multi-region division, routing, and task offloading.

\shen{This framework operates in two stages: \textbf{region division stage} and \textbf{region re-division stage} (See Section~\ref{sec:b}). In the region division stage, the satellite network is divided into multiple uniformly sized regions using a GA-based algorithm. This division is guided by task prediction results generated from a pre-trained Long Short-Term Memory (LSTM) model, ensuring balanced workloads across the network. Based on the obtained network, the routing for the arriving task is initiated at the source satellite within each region (See Section~\ref{sec:routing}). The objective of this process is to identify satellites with sufficient available computational and communication resources. Tasks are then transmitted along selected routes and processed by the satellites located on those routes. To achieve load balancing among satellites within a region, a MA-DDPG-based task splitting and offloading algorithm is employed  (See Section~\ref{sec:taskoff}). Each task is divided into multiple sub-tasks, with each satellite along the route responsible for processing a segment of the task. If a severe workload imbalance is detected between a region and its neighboring regions while the satellite network continues to handle incoming tasks, the region re-division stage is triggered (Stage 2 in \figurename~\ref{fig:framework}). In this stage, the size of the affected region is adjusted using a GA-based algorithm. Further details on the dynamic multi-region division, routing, and task offloading scheme are provided in the following subsections.}

\begin{figure*}[tb!]
    \centering    \includegraphics[width=0.85\textwidth]{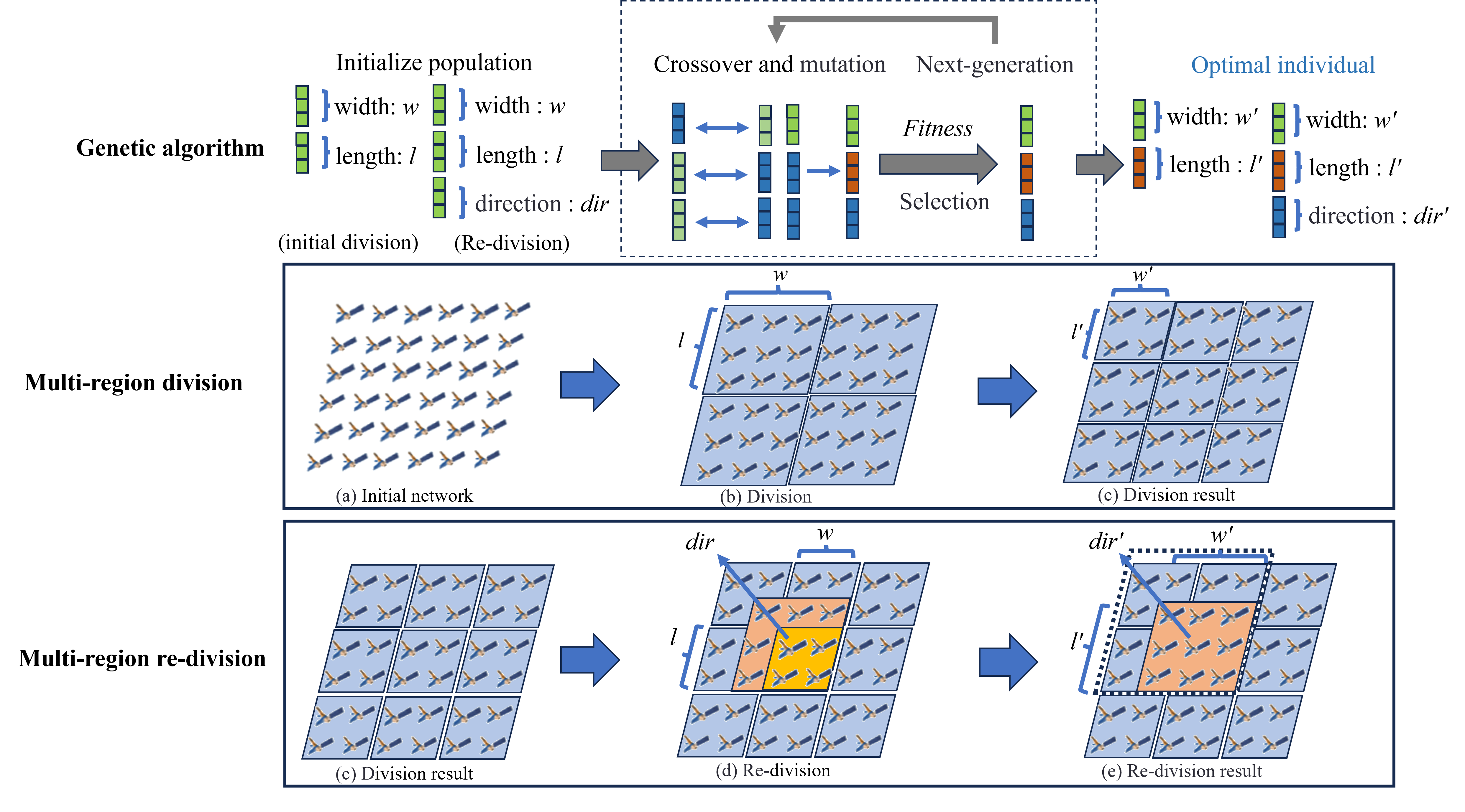}
    \caption{Dynamic multi-region division based on Genetic Algorithm. }
    \label{fig:Region division}
\end{figure*}

\subsection{Dynamic Multi-region Division Algorithm}~\label{sec:b}
To support this multi-region division process, we propose a dynamic multi-region division algorithm that leverages the global optimization capabilities of GA, rooted in the principles of natural selection and evolutionary computation.

Before performing region division, it is essential to predict the tasks originating from source satellites within each candidate region. Simulating the processing of these predicted tasks allows for dynamic adaptation of region sizes to better align with incoming workload demands. To enable this prediction, we employ a pre-trained LSTM-based time series forecasting model. The LSTM architecture is particularly well-suited for this application. I
t effectively captures non-linear patterns in satellite task dynamics, automatically determines the optimal temporal granularity for processing sequential data, and retains long-term dependencies from historical telemetry records.


During the region division phase, we designate $N_{src}$ source satellites, each assigned $M$ tasks, resulting in a total of $N_{src}\times M$ tasks to be allocated. To determine optimized and consistent region dimensions, we develop a GA-based algorithm that generates balanced region division solutions. As illustrated in Fig.~\ref{fig:Region division}, each chromosome is defined by two traits, denoted as $l$ and $w$, representing the length and width of the region, respectively. Corresponding to each chromosome, the entire satellite network is divided into multiple regions of size $l\times w$ (Fig.~\ref{fig:Region division}(a)). The values of 
$l$ and $w$ are within the following ranges:
\begin{equation}
\label{eq:range_l}
K \leq l \leq \sqrt{N}
\end{equation}
\begin{equation}
\label{eq:range_w}
K \leq w \leq \sqrt{N}
\end{equation}

Given that each task can be partitioned into up to $K$ sub-tasks, defining the minimum length and width of each region as $K$ guarantees that all sub-tasks can be executed within the designated area. Subsequently, a routing planning algorithm is executed for each predicted task from the source satellite. Upon task arrival, it is divided into $K$ sub-tasks and offloaded to the satellites along the precomputed route for processing. Detailed descriptions of the routing and task offloading algorithms are provided in Sections~\ref{sec:routing} and \ref{sec:taskoff}. The average task delay $\overline{t}$ and task completion rate $b_{m}$ across all tasks are utilized to compute the fitness value in the GA-based algorithm, which is defined as follows:
\begin{equation}
    \textit{fitness} = \frac{b_m}{\overline{t}}
    \label{eq:fitness}
\end{equation}
Through iterative optimization, an optimal region size $l'\times w'$ for dividing the entire network is determined, as depicted in Fig.~\ref{fig:Region division}(c).

In the region re-division stage, regions experiencing high traffic loads are dynamically resized to mitigate network congestion. This re-division is triggered when a region’s task failure rate exceeds a predefined threshold $\text{THR}_\text{TFR}$, or when its average task delay significantly exceeds neighboring regions. The re-division process is guided by a \shen{GA-based} algorithm to ensure optimal region dimensions and effective load balancing across the network.

As depicted in Fig.~\ref{fig:Region division}, each chromosome is characterized by three traits: $l$ and $w$, and $dir$, where $dir$ denotes the direction of change in region size. When the size of a congested region is adjusted to
$l\times w$ along the direction $dir$, partial overlap with neighboring regions may occur. To address potential resource contention in these overlapping areas, we simulate task processing not only within the resized region but also within its overlapping neighboring regions (see Fig.~\ref{fig:Region division}(e)). For tasks in these regions, routing and task offloading are also planned to obtain the corresponding average task delay and task completion rate, as described in Step 2 of \figurename~\ref{fig:framework}. These performance metrics are then used to compute the fitness value according to Eq.~\ref{eq:fitness}. Through iterative refinement, this process converges toward an optimal re-division size.

\subsection{Adaptive Routing Algorithm}~\label{sec:routing}
Before task allocation, it is essential to determine the optimal route from the source satellite to the destination satellite for the incoming tasks. This ensures efficient utilization of computing and communication resources along the selected route. Due to the multi-hop distances between source and destination satellites and the highly dynamic nature of the satellite network topology, conventional routing algorithms such as Dijkstra’s algorithm can incur significant computational overhead.

\begin{algorithm}
\caption{Adaptive Routing Algorithm}
\label{alg:routingalg}
\begin{algorithmic}[1]
\State \textbf{Input:} Source satellite $S$, destination satellite $D$, network topology $G$
\State \textbf{Output:} Optimal route $R_{opt}$ from $S$ to $D$

\Procedure{RoutePlanning}{$S, D, G$}\label{step:RoutePlanning}
    \State Initialize population $P$ of routes using GA with satellite numbers as chromosomes
    \While{not converged}
        \State Evaluate fitness of each route in $P$ based on Eq.~\ref{eq:evaluateRoute}
        \State Reproduce routes through crossover and mutation.
        \State Update population $P$.
    \EndWhile
    \State \textbf{return} Current optimal route $R_{cur}$ found by GA
\EndProcedure

\State Initialize optimal route $R_{opt} \gets []$
\State $R_{cur} \gets \textit{RoutePlanning}(S, D, G)$
\State Current satellite $C \gets S$
    \While{$C \neq D$}
        \State Next-hop satellite $N \gets \textit{NextSatellite}(R_{cur}, C)$
        \If{\text{PredictedTransmissionDelay}$(C, N)>\textit{Thres}_\textit{d}\label{step:predictDelay}$}
            \State \textbf{Replan Route by:}\label{step:reRoute}
            \State $R_{new} \gets \textit{RoutePlanning}(C, D, G)$
            \State $R_{cur} \gets R_{new}$
        \EndIf
        \State Next-hop satellite $N \gets \textit{NextSatellite}(R_{cur}, C)$
        \State Transmit data from $C$ to $N$
        \State $R_{opt} \gets N$
        \State $C \gets N$
    \EndWhile
\State \Return {$R_{opt}$}

\end{algorithmic}
\end{algorithm}

To address this challenge, we develop a GA-based adaptive routing algorithm as illustrated in Algorithm~\ref{alg:routingalg}. This algorithm consists of two components: a route planning module that constructs a \shen{route between a source satellite and a destination satellite}, and an adaptive routing strategy that dynamically adjusts the planned route if disruptions, such as topology changes, are detected during task execution. When a task arrives at a source satellite, a route is generated to the destination satellite based on the GA. This process is shown in the $RoutePlanning$ function at line~\ref{step:RoutePlanning} of Algorithm~\ref{alg:routingalg}. Within this function, \shen{each GA individual represents a candidate route. } The solution space is explored and refined through genetic operations such as crossover and mutation, followed by the selection of superior individuals based on a fitness evaluation.

During the selection process, the fitness value of each individual is evaluated based on the average resource richness of the satellites along the proposed route. Specifically, for a satellite $k$, its resource richness is expressed by:
\begin{equation}
    u_{k} = c_{k}+r_{k}
    \label{eq:richness}
\end{equation}
where $c_{k}$ and $r_{k}$ denote the percentage of remaining computing and communication resources of satellite $k$. The average resource richness of the satellites along a given route is calculated as:
\begin{equation}
    f(x_i) = \frac{1}{|x_i|} \sum_{k \in x_i} u_k
    \label{eq:evaluateRoute}
\end{equation}
where $|x_i|$ is the number of satellites in the route represented by $x_i$. The acquisition of this resource information is facilitated by the region-based division of the satellite network, as described in Section~\ref{sec:b}, which facilitates efficient sharing of resource status among satellites within each region.

After the route planning, the dynamic nature of the satellite network topology may lead to transmission interruptions during task delivery along the designated route. Such disruptions can occur if the transmission delay between two consecutive satellites exceeds a predefined threshold $Thres_{d}$, shown as line~\ref{step:predictDelay} in Algorithm~\ref{alg:routingalg}. To address this, the proposed algorithm incorporates an adaptive routing mechanism. Specifically, the algorithm continuously monitors the remaining communication resources of the next-hop satellite and predicts the transmission delay for each hop. If a potential connection interruption is detected,
\shen{a new source satellite replan the route as shown in line~\ref{step:reRoute} of Algorithm~\ref{alg:routingalg}.}
This adaptive strategy enhances the robustness of task transmission and ensures reliable processing under dynamic network conditions.

The previous discussion addressed the route planning problem between source and destination satellites within the same region. However, the source satellite lacks sufficient information about resource availability in other regions, making it infeasible to generate a complete route in a single step in cross-region scenarios. Therefore, the route generation process is divided into two phases, both executed in accordance with Algorithm~\ref{alg:routingalg}. The first phase involves routing from the starting satellite to a boundary satellite within its region. \shen{The second phase then establishes the route from this boundary satellite to the destination satellite in the neighboring region.} This hierarchical routing strategy leverages the critical role of boundary satellites, which have visibility into both intra-regional resource states and inter-regional resource availability in adjacent regions. By functioning as gateways, boundary satellites enable more informed routing decisions, balancing computing and communication resource usages across regions and ensuring efficient route planning.

\shen{The adaptive routing scheme requires iterating over a population of size $P$ for $G$ generations, where each candidate route has up to $L$ hops. Each generation performs selection, crossover, mutation and fitness evaluation for $P$ individuals. Since fitness (cost) is computed in time proportional to the route length $L$, the dominant cost per generation is $O(P \cdot L)$. Over G generations, the time complexity is therefore on the order of $O(G \cdot P \cdot L)$. Moreover, the total complexity for the adaptive routing algorithm is also dependent on the number of re-planning instances. In the worst case, if the route is adjusted for each hop, the time complexity for the entire adjustment phase would also be $O(G \cdot P \cdot L^{2})$.}

\subsection{Task Splitting and Offloading Algorithm}~\label{sec:taskoff}
We propose a task-splitting and offloading algorithm based on the MA-DDPG framework, tailored for satellite task management. The primary objective is to achieve load balancing and to alleviate excessive resource consumption \shen{in the vicinity of the source satellite.}

MA-DDPG extends the DDPG algorithm by incorporating multi-agent coordination, which makes it particularly suitable for distributed satellite networks. In this context, MA-DDPG is employed to determine the optimal sub-task division ratios and to select appropriate satellites for executing the sub-tasks. Each source satellite operates as an independent agent for deciding the offloading \shen{actions}. This approach ensures that resource utilization is balanced across the network, reducing congestion near the source satellite, and enhances overall system efficiency.


The offloading process is constructed as a Markov Decision Process defined as $\mathcal{M}=(\mathcal{S},\mathcal{A},\mathcal{R})$, where the elements $\mathcal{S}$, $\mathcal{A}$, $\mathcal{R}$ represent the state space, action space, and reward function, respectively. Each element of the MDP is defined as below:


1) State $\mathcal{S}$: The agent acquires state information from the environment at the beginning of every time slot $t$. Each agent $i$ observes its state information as follows:
\begin{equation}
\begin{split}
    s_{i}(t) = \left\{ \bm{c_{k}}, \bm{r_{k}}, \bm{o_{k}},
    q
    \right\}
\end{split}
\end{equation}
where $\bm{c_{k}}$ = $\{c_{1},c_{2},\cdots, c_{K}\}$ and $\bm{r_{k}}$ = $\{r_{1},r_{2},\cdots, r_{K}\}$ denote the the percentage of remaining computation resources and communication resources of the first $K$ satellite, $\bm{o_{k}}$ = $\{o_{1},o_{2},\cdots, o_{K}\}$ indicates \shen{the number of times the first $K$ satellites of the route has already been selected by other agents' assigned tasks, aiming to prevent certain satellites from being over-allocated tasks}, and $q$ denotes task size.

2) Action $\mathcal{A}$: Based on the network's current state, each agent determines and executes its corresponding actions. Given the distributed multi-agent architecture, each agent operates independently with its own set of actions. The action space is defined as follows:
\begin{equation}
    a_{i}(t) = \left\{ \bm{p_{k}^{task}}, \bm{p_{k}^{comp}}, \bm{p_{k}^{comm}} \right\}
\end{equation}
where $\bm{p_{k}^{task}} = \{p_{1}^{task},p_{2}^{task},\cdots, p_{K}^{task}\}$ denotes the \shen{division ratio} of $K$ sub-task. \shen{The $k$-th sub-task is allocated to the corresponding $k$-th satellite on the route (determined by the routing algorithm) for execution.} Similarly, $\bm{p_{k}^{comp}}$ = $\{p_{1}^{comp},p_{2}^{comp},\cdots, p_{K}^{comp}\}$ and $\bm{p_{k}^{comm}}$ = $\{p_{1}^{comm},p_{2}^{comm},\cdots, p_{K}^{comm}\}$ represent the proportions of computational and communication resources allocated to each of the allocated to $K$ sub-task, respectively.

3) Reward $\mathcal{R}$: After all agents have made their decisions, the reward value is calculated according to Eq.~\ref{eq:reward}. This reward is based on the average delay of completed tasks per agent and the overall task completion rate.
The reward function is defined as follows:
\begin{equation}
    r(t) = \Re-J(t)-\rho \sigma^2
    \label{eq:reward}
\end{equation}
\shen{where $\Re$ is a constant that ensures the reward tends to be positive. $J(t)$ represents the optimization objective (See Eq.~\ref{eq:obj}) in time slot $t$. $\rho$ is the scaling coefficient. $\sigma^2$ denotes the variance in the resource availability of satellites along the task route. It is incorporated into the task allocation strategy to prevent uneven sub-task assignments that might excessively burden certain satellites.}

\begin{algorithm}[tb]
\small
    \renewcommand{\algorithmicrequire}{\textbf{Input:}}
    \renewcommand{\algorithmicensure}{\textbf{Output:}}
    \caption{MA-DDPG-based Task Offloading Scheme}
    \label{alg:task offloading}
    \begin{algorithmic}[1]
        \Require Source satellites $v_{i} \in V$ and tasks to be processed on each source satellite $q_{m}^{i}\in Q$, the well-trained policy net $\pi$.
        \Ensure The task offloading result $scheme$.
        \State Initialize $scheme \leftarrow []$
        \For{each task $m$\label{algline:b1}}
        \State route set $routes
        \leftarrow \emptyset,$ task set $tasks
        \leftarrow \emptyset$
        \For{each source satellite $v_{i}$\label{algline:b2}}
        \State Select task $q_{m}^{i}$ from available tasks $\{q_{1}^{i}, q_{2}^{i}, \cdots, q_{M}^{i}\}$ for satellite $v_{i}$
        \State $route_{m}^{i}\leftarrow Routing(v_{i}, q_{m}^{i})\label{algline:b3}$ 
        \State $routes \leftarrow routes \cup \{route_{m}^{i}\}\label{algline:b4}$
        \State $tasks \leftarrow tasks \cup \{q_{m}^{i}\}$
        \EndFor
        \For{each source satellite $v_{i}$}
        \State \textbf{Obtain state $s_{i}$:}
        \For{each satellite $v_{k} \in routes[i]$}
        \State $s_{i} \gets c_{k},r_{k},o_{k}$
        \EndFor
        \State \textbf{Obtain action $a_{i}$:}
        \State $a_{i} \gets \pi_{i}(s_{i}|\theta_{\pi})$
        \State $scheme \gets a_{i}$
        \EndFor
        \State Execute each task offloading action $a_{i}$ for $v_{i}$.
        \State Update network resource status.\label{algline:b5}
        \EndFor
    \end{algorithmic}
\end{algorithm}

In our algorithm, MA-DDPG adopts a decentralized actor–critic framework to address multi-agent reinforcement learning challenges under partial observability. Each agent \(i\) maintains two neural networks:

\textbf{Actor (Policy Network)}: The actor \(\mu_{\theta_i}\) maps the agent's local observations \(o_i\) to a deterministic action \(a_i\), enabling decentralized execution. The policy gradient update is given by:
    \begin{equation}
        \nabla_{\theta_i} J(\theta_i) = \mathbb{E}_{o_i \sim \mathcal{D}} \left[ \nabla_{\theta_i} \mu_{\theta_i}(o_i) \cdot \nabla_{a_i} Q_i^{\phi_i}(s,\mathbf{a})\big|_{a_i=\mu_{\theta_i}(o_i)} \right]
    \end{equation}
    where \(s\) denotes the global state and \(\mathbf{a} = (a_1,...,a_N)\) represents joint actions. Target network parameters \(\theta_i'\) are updated via Polyak averaging: 
    \begin{equation}
    \theta_i' \gets \tau \theta_i + (1 - \tau) \theta_i'
\end{equation}
where \(\tau \in (0,1)\) is the \shen{soft update rate} controlling the update rate of target networks.
    
\textbf{Critic (Value Network)}: The critic \(Q_i^{\phi_i}\) estimates action values using \emph{centralized training} with global information, minimizing the temporal difference (TD) error:
    \begin{equation}
        \mathcal{L}(\phi_i) = \mathbb{E}_{(s,\mathbf{a},r_i,s') \sim \mathcal{D}} \left[ \left( Q_i^{\phi_i}(s,\mathbf{a}) - \left( r_i + \gamma Q_i^{\phi_i'}(s',\mathbf{a}') \right) \right)^2 \right]
    \end{equation}
where \(\mathbf{a}' = (\mu_{\theta_1'}(o_1'),...,\mu_{\theta_N'}(o_N'))\) and \(\phi_i'\) denotes target critic parameters updated by: 
\begin{equation}
    \phi_i' \gets \tau \phi_i + (1 - \tau) \phi_i'
\end{equation}

This multi-agent distributed architecture follows the centralized training with decentralized execution paradigm, where critics leverage global state information to guide policy updates while actors operate using only local observations during deployment.

    




    

Algorithm~\ref{alg:task offloading} describes the process of task offloading, which incorporates a well-trained policy network $\pi_{i}(s_{i}|\theta_{\pi})$ to make offloading decisions. As input, we select a subset of satellites from the network to serve as source satellites. These source satellites act as agents, each containing 
tasks that need to be allocated. As shown in lines~\ref{algline:b1} and ~\ref{algline:b2} of Algorithm~\ref{alg:task offloading}, each agent $v_{i}$ assigns a task $q_{m}^{i}$. We use the algorithm in Section C to plan route for each task, depicted in line~\ref{algline:b3} of Algorithm~\ref{alg:task offloading}. To determine the intersection of the planned routes for each task with the routes of other tasks (i.e., $o$ in state 
$s$), we collect the execution routes of tasks on each source satellite into the set $routes$ as shown in line~\ref{algline:b4} of Algorithm~\ref{alg:task offloading}. Then, we determine the state 
$s_{i}$ of each agent $v_{i}$ using the $routes$ and $tasks$ sets, and obtain the offloading scheme based on the well-trained policy network. After each round of task allocation, we update the resource status of the satellites, as shown in line~\ref{algline:b5} of Algorithm~\ref{alg:task offloading}.

\shen{The time computational complexity of the MA-DDPG-based task splitting and offloading algorithm is primarily determined by the number of ground stations $N_G$, tasks per batch $M$, number of batches $B$, and the architecture of the Actor neural network. Assuming an Actor network with $L_A$ layers and $H_\ell$ neurons per layer, the complexity of a single forward pass is $O(L_A H_\ell^2)$. Since each task across all agents and batches requires such evaluations, the total time complexity is $O(N_G \cdot M \cdot B \cdot L_A H_\ell^2)$.
}

{
\section{Experiments}
\subsection{Experimental Setup}

\begin{table}[tb!]
 \caption{Main experimental parameters}
 \label{tab:params}
 \centering
 \scalebox{0.8}{
\begin{tabular}{ll}
   \hline
   Parameter  & Value \\
   \hline 

        Number of satellites in the constellation $N_{S}$ & 1600/3600\\

        Number of ground station $N_{G}$ & 100\\

        Number of tasks per batch on each ground station
 \textit{M} &10\\

        Number of task batches \textit{B} & 10\\

        Task splitting number
 \textit{K} & 5\\
        Memory replay buffer size $R$ & 1000\\
        Mini-batch sample size $N$ & 64\\
        Discount factor $\gamma$ & 0.98 \\
        Ornstein-Uhlenbeck noise $\sigma$ &0.5 \\
        Soft update factor $\tau$ & 0.01\\
        Actor network learning rate \ &0.001 \\
        Critic network learning rate \ &0.001 \\
        \hline
  \end{tabular}
  }
\end{table}

We conduct extensive numerical experiments to validate the performance of our proposed system, \sysname. The experiments are run on a CPU-based server equipped with 16~GB 4800-MHz DDR5, 2.50-GHz Intel Core i5. We develop a simulation environment for a LEO satellite constellation using a satellite network routing simulator~\cite{ccr20}, which models the SpaceX Starlink network. We consider two network scales: 1600 satellites (40$\times$40) and 3600 satellites (60$\times$60).  For task allocation, the number of task-receiving source satellites is set to $S$=5, with each source satellite handling $M$=10 tasks per batch. This results in a total of $S\times M\times B$ tasks across all simulation batches. The main parameters used in our experiments are summarized in Table~\ref{tab:params}.

The proposed framework, \sysname, consists of three main modules: region division, route planning, and task allocation. The effectiveness of the dynamic region division algorithm is demonstrated through comparisons with two baselines: static region division and no division. In addition, the framework incorporates routing, task splitting, and task offloading modules. We compare the algorithms used in these modules with alternative baseline methods to assess their relative performance. We evaluate the system using three key metrics: task completion rate, average delay of completed tasks, and average energy consumption per task. The results are summarized as follows:

\subsection{Dynamic Multi-region Division Performance}

In this subsection, we evaluate the effectiveness of the dynamic multi-region division module in our proposed framework, \textit{\textbf{\sysname}}, by comparing it against the following three baseline methods:

\begin{itemize}
\item \textbf{QuadTree Region Division (\textit{QTRD})} is based on a quadtree division approach, recursively dividing each region into four sub-blocks until the load is balanced across all sub-blocks~\cite{Jiang2021Aware}.
\item \textbf{Grid-based Region Division (\textit{GRD})} divides the satellite network into equal-sized square regions and determines the optimal division granularity through enumerating different region sizes~\cite{li2023multi}.
\item \textbf{No Region Division (\textit{NRD})} indicates that no region division is applied; instead, tasks are processed by any available neighboring satellites.
\end{itemize}

\figurename~\ref{fig:RD40_result} and \figurename~\ref{fig:RD60_result} present the performance of different region division methods under varying network scales. For a network size of 1600 satellites, the task completion rate gradually decreases as the average task size increases. The results demonstrate that $\textit{\sysname}$ significantly outperforms QuadTree Region Division (\textit{QTRD}), Grid-based Region Division (\textit{GRD}), and No Region Division (\textit{NRD}). This improvement is attributed to \textit{\sysname}'s dynamic task partitioning across regions, which mitigates network congestion. During congestion, \textit{\sysname} adaptively adjusts region sizes to alleviate bottlenecks, thereby improving the task completion rate. Specifically, \textit{\sysname} achieves task completion rates approximately 1.11\%, 3.51\%, and 5.78\% higher than \textit{QTRD}, \textit{GRD}, and \textit{NRD}, respectively. Although larger task sizes increase time and energy demands due to limited satellite resources, \textit{\sysname} maintains relatively lower delay and energy consumption. This is achieved through adaptive region re-division, which reduces task offloading overlaps from different source satellites to the same satellite, thereby optimizing regional resource utilization. In the 40$\times$40 satellite network, \textit{\sysname} achieves an average delay reduction of 129.3 ms, 216.1 ms, and 330.5 ms, and energy savings of 0.038 J, 0.085 J, and 0.165 J compared to \textit{QTRD}, \textit{GRD}, and \textit{NRD}, respectively.

On the other hand, in the larger 3600-satellite (60$\times$60) network, \textit{\sysname} continues to deliver strong performance. Compared to the 40$\times$40 network, it achieves further improvements in average task delay, task completion rate, and energy consumption. These enhancements arise from the ability to apply finer-grained region division in larger networks, which leads to better load balancing among satellites within each region. For the subsequent experiments, all methods adopt dynamic multi-region division to ensure a fair and consistent comparison.

\begin{figure*}
        \centering
	\begin{minipage}[b]{0.8\columnwidth}
		\centering
		\includegraphics[width=\columnwidth]{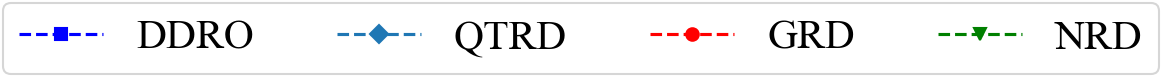}
	\end{minipage}
    \vspace{-0.5cm}
\end{figure*}
  \begin{figure*}[tb]
	\centering
    \begin{minipage}[b]{.5\columnwidth}
		\centering		\includegraphics[width=\columnwidth]{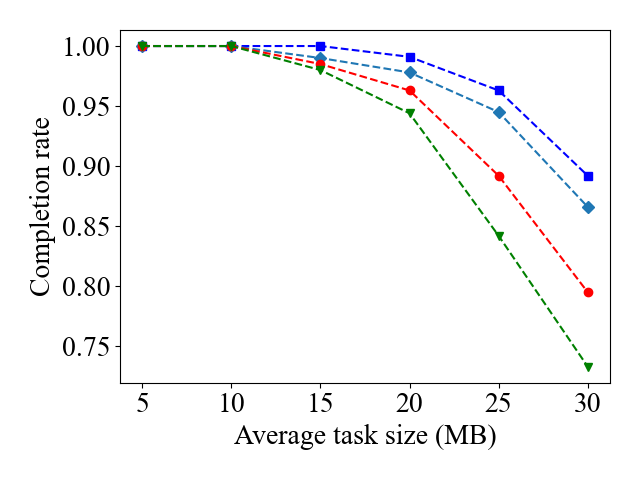}
		\subcaption{Task completion rate
         }\label{fig:time_r1}
	\end{minipage}
    \begin{minipage}[b]{.5\columnwidth}
		\centering		\includegraphics[width=\columnwidth]{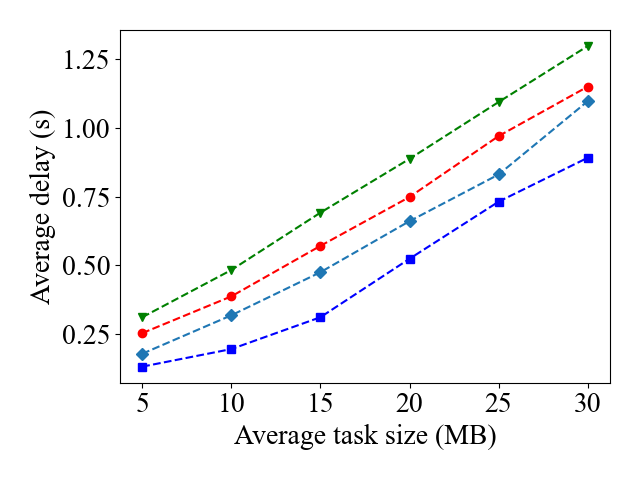}
		\subcaption{Total average delay }
	\end{minipage}
    \begin{minipage}[b]{.5\columnwidth}
		\centering
		\includegraphics[width=\columnwidth]{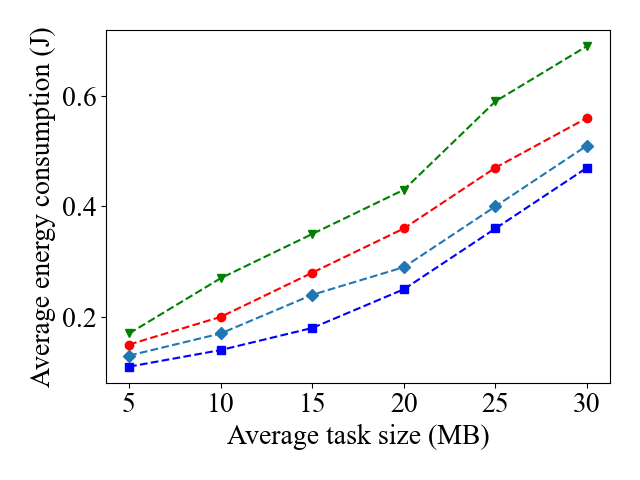}
		\subcaption{Average energy consumption
         }
	\end{minipage}
	\caption{Performance comparison of different region division methods when the network size is 1600.}
	\label{fig:RD40_result}
 
 \end{figure*}

 \begin{figure*}
        \centering
	\begin{minipage}[b]{0.8\columnwidth}
		\centering
		\includegraphics[width=\columnwidth]{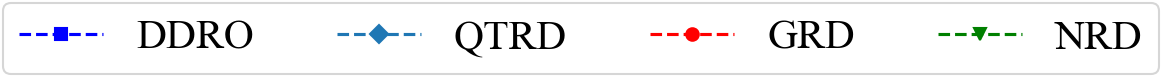}
	\end{minipage}
    \vspace{-0.5cm}
\end{figure*}
  \begin{figure*}[tb]
	\centering
    \begin{minipage}[b]{.5\columnwidth}
		\centering		\includegraphics[width=\columnwidth]{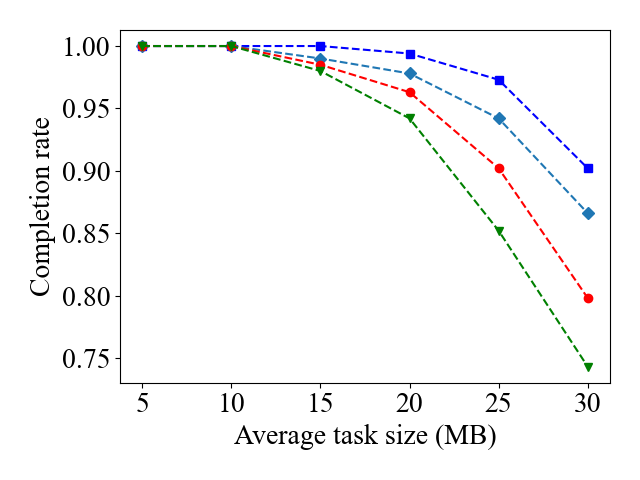}
		\subcaption{Task completion rate
         }
	\end{minipage}
    \begin{minipage}[b]{.5\columnwidth}
		\centering		\includegraphics[width=\columnwidth]{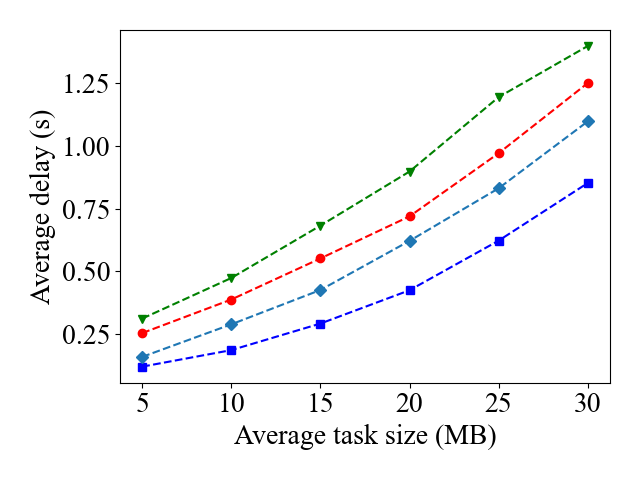}
		\subcaption{Total average delay }
	\end{minipage}
    \begin{minipage}[b]{.5\columnwidth}
		\centering
		\includegraphics[width=\columnwidth]{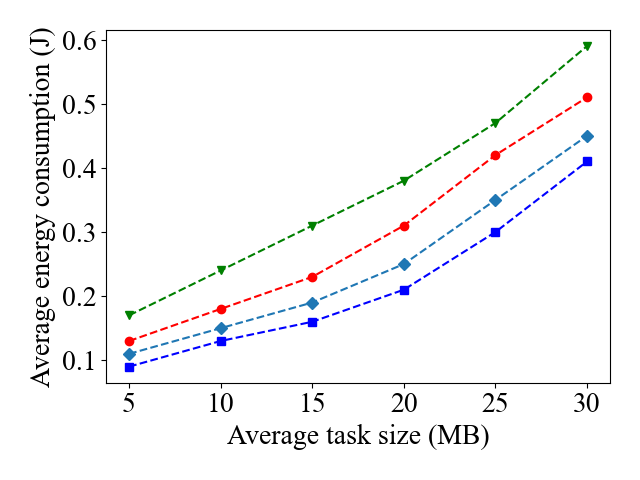}
		\subcaption{Average energy consumption
         }
	\end{minipage}
	\caption{Performance comparison of different region division methods when the network size is 3600.}
	\label{fig:RD60_result}
 
 \end{figure*}

\subsection{Routing Performance}

In this subsection, we evaluate the effectiveness of the route planning module in \textit{\textbf{\sysname}} by comparing it with the following three baseline methods:

\begin{itemize}
\item  \textbf{Centralized Route Management \shen{Planning} (\textit{CRP})} leverages SDN technology and Dijkstra’s algorithm to generate end-to-end routing plans based on the current satellite network topology~\cite{QiFGCS22}.
\item  \textbf{Opportunistic Resource Priority (\textit{ORP})} constructs routes by selecting the satellite with the highest residual resources from the available options in each step~\cite{jiao2024clustered}.
\item  \textbf{Random Step Hop (\textit{RSH})} generates routes by randomly and independently selecting the next-hop satellite from the available candidates at each step.
\end{itemize}

\figurename~\ref{fig:route_result} summarizes the performance of various route planning methods when the network size is 1600. The results indicate that \textit{\sysname} outperforms \textit{CRP}, \textit{ORP}, and \textit{RSH}. This superior performance of \textit{\sysname} is attributed to using a comprehensive evaluation metric that considers the average residual resources across all satellites along the route. \textit{\sysname} effectively addresses a key limitation of \textit{ORP}, which often falls into local optima by focusing solely on immediate resource availability and neglecting the broader network context.

The routes generated by \textit{\sysname} are characterized by high levels of residual resources across the satellite network, providing robust support for both task processing and data transmission. \textit{\sysname} achieves an average delay reduction of 76.8 ms, 183.6 ms, and 314.3 ms compared to \textit{CRP}, \textit{ORP}, and \textit{RSH}, respectively.
Although larger task sizes increase the risk of transmission interruptions due to longer delays between adjacent satellites, \textit{\sysname} maintains a relatively higher task completion rate. This is achieved through dynamic route re-planning when excessive transmission delays are detected. Specifically, \textit{\sysname} improves the task completion rate by approximately 6.35\%, 3.61\%, and 10.76\% over \textit{CRP}, \textit{ORP}, and \textit{RSH}, respectively.

\begin{figure}
        \centering
	\begin{minipage}[b]{0.75\columnwidth}
		\centering
		\includegraphics[width=\columnwidth]{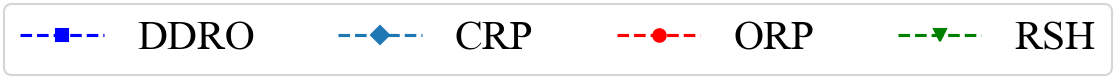}
	\end{minipage}
    \vspace{-0.3cm}
\end{figure}
  \begin{figure}[tb]
	\centering
    \begin{minipage}[b]{.48\columnwidth}
		\centering		\includegraphics[width=\columnwidth]{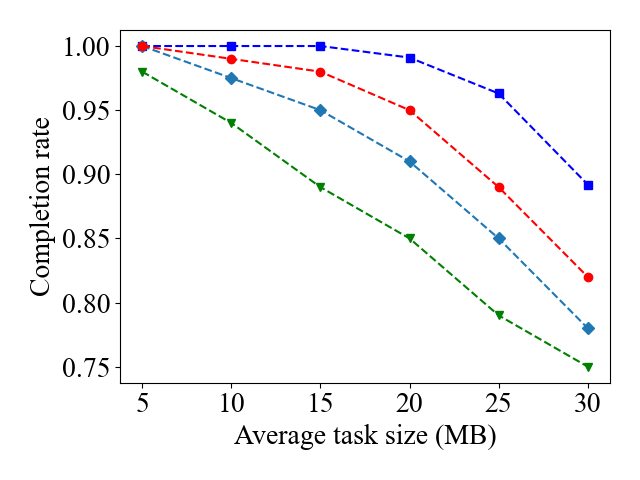}
		\subcaption{Task completion rate
         }
	\end{minipage}
    \begin{minipage}[b]{.48\columnwidth}
		\centering		\includegraphics[width=\columnwidth]{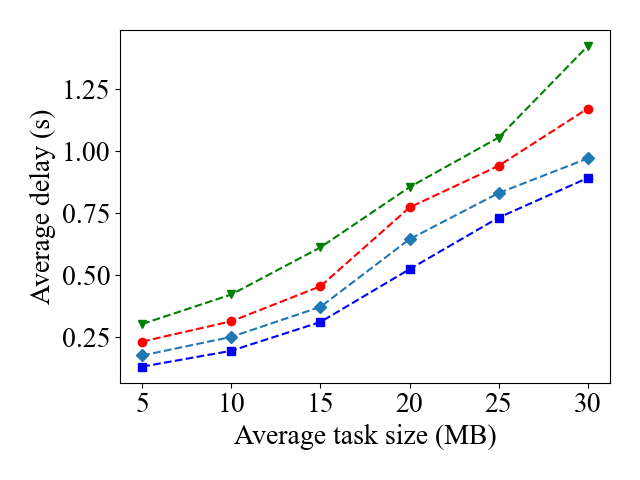}
		\subcaption{Total average delay }\label{fig:tcr_r1}
	\end{minipage}
	\caption{Performance comparison of different routing methods when the network size is 1600.}
	\label{fig:route_result}
 
 \end{figure}

\subsection{Task Splitting and Offloading Performance}



\begin{figure*}
        \centering
	\begin{minipage}[b]{1.0\columnwidth}
		\centering
		\includegraphics[width=\columnwidth]{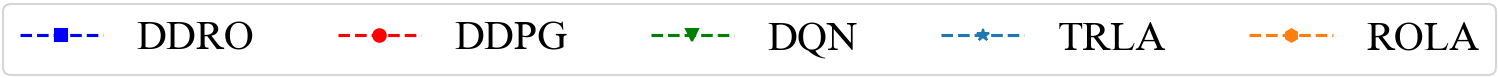}
	\end{minipage}
    \vspace{-0.5cm}
\end{figure*}
  \begin{figure*}[tb]
	\centering
    
    \begin{minipage}[b]{.5\columnwidth}
		\centering		\includegraphics[width=\columnwidth]{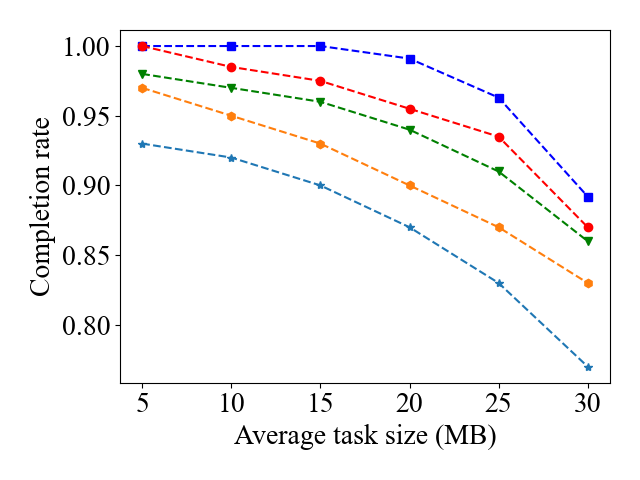}
		\subcaption{Task completion rate
         }
	\end{minipage}
    \begin{minipage}[b]{.5\columnwidth}
		\centering		\includegraphics[width=\columnwidth]{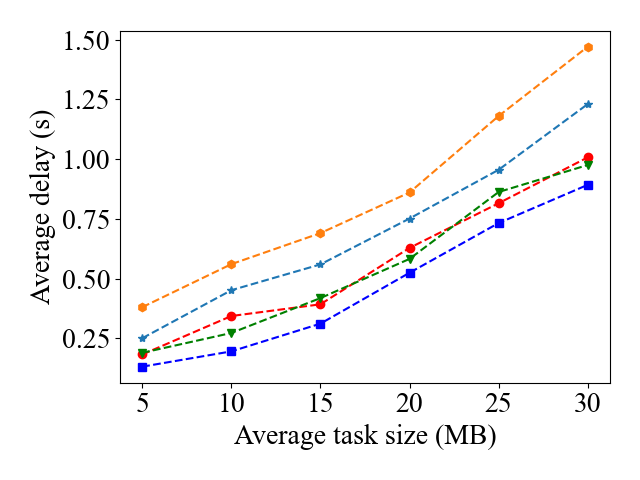}
		\subcaption{Total average delay}
	\end{minipage}
    \begin{minipage}[b]{.5\columnwidth}
		\centering
		\includegraphics[width=\columnwidth]{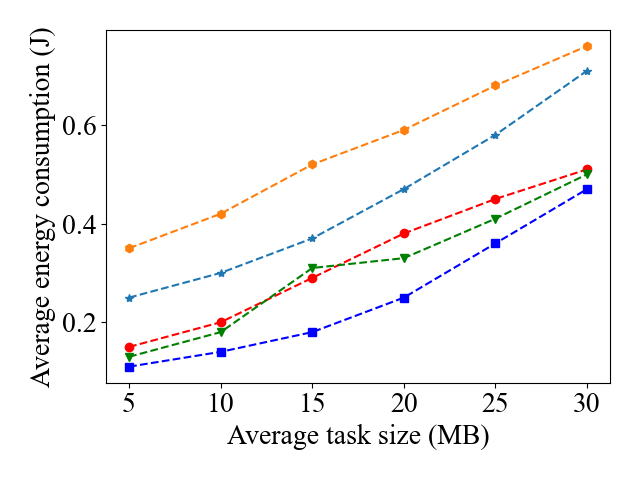}
		\subcaption{Average energy consumption
         }
	\end{minipage}
    \begin{minipage}[b]{.5\columnwidth}
		\centering
		\includegraphics[width=\columnwidth]{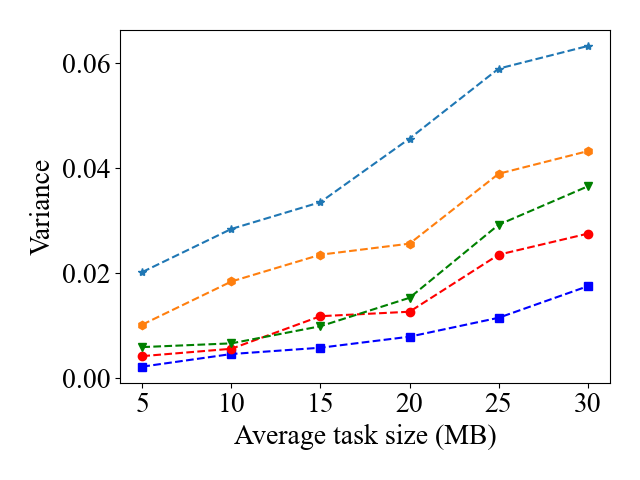}
		\subcaption{Variance
         }
	\end{minipage}
	\caption{Performance comparison of different task splitting scheme methods when the network size is 1600.}
	\label{fig:result4}
 
 \end{figure*}

 \begin{figure*}
        \centering
	\begin{minipage}[b]{1.0\columnwidth}
		\centering
		\includegraphics[width=\columnwidth]{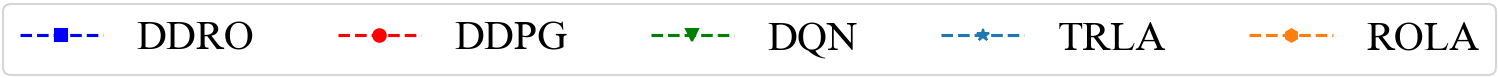}
	\end{minipage}
    \vspace{-0.5cm}
\end{figure*}
  \begin{figure*}[tb]
	\centering
    \begin{minipage}[b]{.5\columnwidth}
		\centering		\includegraphics[width=\columnwidth]{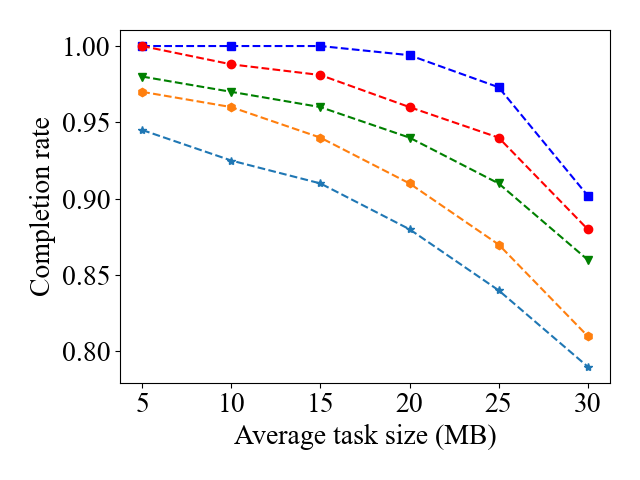}
		\subcaption{Task completion rate
         }
	\end{minipage}
    \begin{minipage}[b]{.5\columnwidth}
		\centering		\includegraphics[width=\columnwidth]{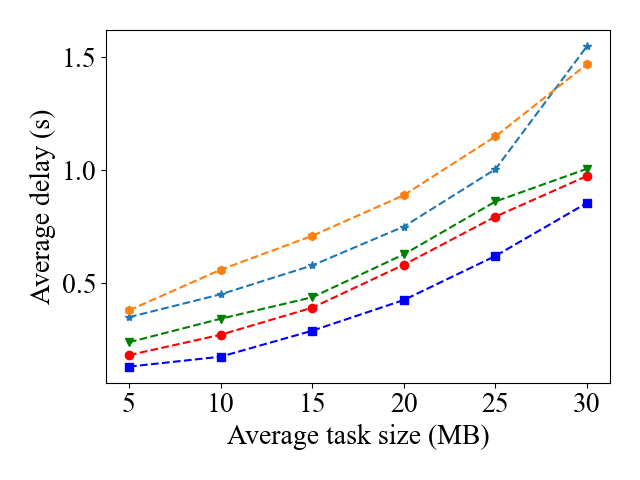}
		\subcaption{Total average delay }
	\end{minipage}
    \begin{minipage}[b]{.5\columnwidth}
		\centering
		\includegraphics[width=\columnwidth]{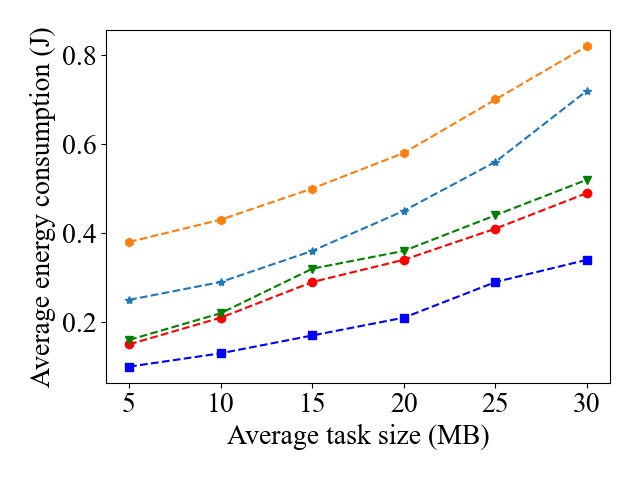}
		\subcaption{Average energy consumption
         }
	\end{minipage}
    \begin{minipage}[b]{.5\columnwidth}
		\centering
		\includegraphics[width=\columnwidth]{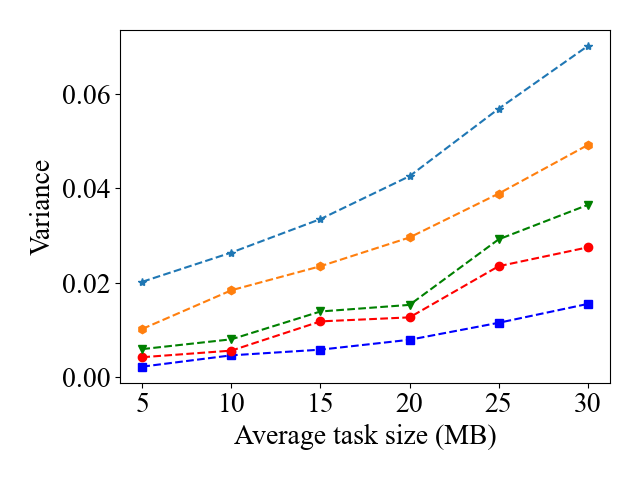}
		\subcaption{Variance
         }
	\end{minipage}
	\caption{Performance comparison of different task splitting scheme methods when the network size is 3600.}
	\label{fig:result5}
 
 \end{figure*}

In this subsection, we evaluate the effectiveness of the task splitting and offloading module in our proposed \textit{\textbf{\sysname}} by comparing it against the following four baseline methods:

\begin{itemize}
\item  \textbf{Deep Deterministic Policy Gradient (\textit{DDPG})} is a DDPG-based task offloading method designed to minimize task drop rate and delay~\cite{zhang2023satellite}.
\item \textbf{Deep Q-Network (\textit{DQN})} is a DQN-based task offloading method that also aims to reduce task drop rate and delay~\cite{ZhengAIoT24}.
\item \textbf{Transmission Loss Reduction Allocation (\textit{TRLA})}: is a method that mitigates task transmission loss by fully utilizing the current satellite's computing resources, transferring remaining tasks to the next satellite only when the current one is fully occupied~\cite{ZhangTNSE23}.
\item  \textbf{Random Offloading Location Assignment (\textit{ROLA})} is a method that randomly splits tasks and allocates them to candidate satellites. 
\end{itemize}

\figurename~\ref{fig:result4} shows the performance of different task segmentation and offloading methods when the network size is 1600. \figurename~\ref{fig:result4}(a) illustrates the task completion rates under different task sizes. As shown, \textit{\sysname} consistently maintains high performance, even under large task size conditions. This is attributed to its use of a multi-agent reinforcement learning algorithm for task offloading. In particular, \textit{\sysname} effectively avoids overloading satellites that are already burdened with tasks, thereby reducing the risk of task drops due to excessive load. Specifically, \textit{\sysname} improves the task completion rate by 2.11\%, 3.76\%, 10.43\%, and 6.61\% compared to \textit{DDPG}, \textit{DQN}, \textit{TRLA}, and \textit{ROLA}, respectively.

\figurename~\ref{fig:result4}(b) and \figurename~\ref{fig:result4}(c) show the average delay and energy consumption per task, respectively. Among the evaluated methods, \textit{\sysname}consistently achieves lower delay and energy consumption by
optimizing both task delay and completion rate during the task offloading process. Through efficient task allocation,  \textit{\sysname} optimizes satellite resource utilization along the route, thereby reducing both latency and energy usage. In contrast, \textit{DDPG} and \textit{DQN} lack the ability to observe the behavior of other agents, which can result in multiple source satellites offloading a large number of tasks to the same satellite. \shen{This uncoordinated behavior can overload satellites that initially have sufficient resources, leading to increased delays and higher energy consumption.} \shen{\textit{TRLA} prioritizes the use of computing resources on the current satellite, only offloading to the next satellite once the local resources are fully utilized}. This approach quickly depletes the resources of the current satellite and neighboring satellites, resulting in a significant resource imbalance. Meanwhile, \textit{ROLA}, which selects actions randomly, suffers from substantial delays and elevated energy consumption due to its lack of a systematic decision-making strategy. Overall, \textit{\sysname} reduces the average delay by 97.3 ms, 86.6 ms, 235.1 ms, and 393.2 ms, and the average energy consumption by 0.078 J, 0.058 J, 0.195 J, and 0.306 J, compared to \textit{DDPG}, \textit{DQN}, \textit{TRLA}, and \textit{ROLA}, respectively.


\figurename~\ref{fig:result4}(d) highlights the maximum resource usage variance across different methods. A lower variance in satellite resource usage indicates a more effective load-balancing strategy across the network. The experimental results show that \textit{\sysname} achieves the lowest variance, demonstrating the most robust load-balancing capability among all evaluated algorithms.

\figurename~\ref{fig:result5} illustrates the performance of different task segmentation and offloading methods in a network with 3600 satellites. As the network size increases, resource disparities among satellites become more pronounced. Ineffective task allocation in such environments can lead to overloading of certain satellites while others remain underutilized, ultimately degrading overall system performance. The consistent performance trends observed across different task allocation methods in the larger 1600-satellite network are attributed to the MA-DDPG foundation of \textit{\sysname}. The proposed framework maintains a balanced exploration–exploitation trade-off and leverages adaptive learning to continuously refine strategies based on environmental feedback. Furthermore, the route planning phase proactively selects task-capable satellites along optimized routes, mitigating the complexity of large-scale search operations. These design features enable \textit{\sysname} to sustain scalability and efficiency in expansive network environments. In addition, as the network scale increases, the effectiveness of \textit{\sysname}'s region division strategy is further enhanced, leading to reduced task delay and improved task completion rates.


\section{Conclusion}
In this paper, we propose a dynamic multi-region division framework, \textit{\sysname}, designed to minimize task delay while balancing heterogeneous resource utilization. The framework divides the entire satellite network into multiple regions, enabling distributed task execution across these regions. Tasks collected by satellites are segmented accordingly, and a joint routing and task offloading algorithm determines the optimal offloading strategy. To further enhance performance, the system continuously monitors regional load conditions and dynamically adjusts the size of high-load regions to alleviate traffic congestion. Experimental results demonstrate that \textit{\sysname} outperforms existing methods across a range of metrics and network scales. Specifically, our framework effectively reduces task execution and transmission delays, lowers energy consumption, and improves task completion rates, all while ensuring efficient load balancing in satellite resource allocation.

\bibliographystyle{ieeetr}
\bibliography{ref.bib}

\section*{Bibliographies}
\vskip -2\baselineskip plus -1fil
\begin{IEEEbiographynophoto}{Zixuan Song} received the B.E. degree from Hubei University, Wuhan, China, in 2020. He is currently pursuing the master's degree with the School of Computer Science and Artificial Intelligence, Wuhan University of Technology, Wuhan, China. His primary research interests focus on satellite communication.
\end{IEEEbiographynophoto}

\vskip -2\baselineskip plus -1fil

\begin{IEEEbiographynophoto}{Zhishu Shen}
received the B.E. degree from the School of Information Engineering at the Wuhan University of Technology, Wuhan, China, in 2009, and the M.E. and Ph.D. degrees in Electrical and Electronic Engineering and Computer Science from Nagoya University, Japan, in 2012 and 2015, respectively. He is currently an Associate Professor in the School of Computer Science and Artificial Intelligence, Wuhan University of Technology. From 2016 to 2021, he was a research engineer of KDDI Research, Inc., Japan. His major interests include network design and optimization, data learning, edge computing and the Internet of Things.
\end{IEEEbiographynophoto}

\vskip -2\baselineskip plus -1fil

\begin{IEEEbiographynophoto}{Xiaoyu Zheng}
received the B.E. degree from Wuhan University of Technology, Wuhan, China, in 2024. He is currently pursuing the master's degree with the School of Computer Science and Artificial Intelligence, Wuhan University of Technology, Wuhan, China. His major research interests focus on network optimization including satellite network and smart-grid network.
\end{IEEEbiographynophoto}

\vskip -2\baselineskip plus -1fil

\begin{IEEEbiographynophoto}{Qiushi Zheng}received the B.E. degree in automation (information and control system) from Beijing Information Science and Technology University, China, in 2015. He received the M.E. degree in Engineering Science (Electrical and Electronic) and the Ph.D degree in Information Communication Technology from the Swinburne University of Technology, Australia, in 2017 and 2023. He is currently a research Fellow at Swinburne University of Technology, Melbourne, Australia. His research interests include satellite-aerial-ground integrated networks, edge computing and the Internet of Things.
\end{IEEEbiographynophoto}

\vskip -2\baselineskip plus -1fil

\begin{IEEEbiographynophoto}
{Zheng Lei} is the inaugural Professor of Aviation and Chair of the Department of Aviation at Swinburne University of Technology, Australia.  Zheng holds a PhD in Aviation from the University of Surrey in 2006.  He has research interests in aviation business models, aviation market analysis, aviation regulation and policy analysis, and uncrewed aircraft systems.  A distinctive feature of his research is its real-world impact and industry focus.  He is also a Board member of Aviation Aerospace Australia. 
\end{IEEEbiographynophoto}

\vskip -2\baselineskip plus -1fil

\begin{IEEEbiographynophoto}{Jiong Jin} received the B.E. degree with First Class Honours in Computer Engineering from Nanyang Technological University, Singapore, in 2006, and the Ph.D. degree in Electrical and Electronic Engineering from the University of Melbourne, Australia, in 2011. He is currently a Full Professor in the School of Engineering, Swinburne University of Technology, Melbourne, Australia. His research interests include network design and optimization, edge computing and intelligence, robotics and automation, and cyber-physical systems and Internet of Things as well as their applications in smart manufacturing, smart transportation and smart cities. He was recognized as an Honourable Mention in the AI 2000 Most Influential Scholars List in IoT (2021 and 2022). He is currently an Associate Editor of IEEE Transactions on Industrial Informatics and IEEE Transactions on Network Science and Engineering.
\end{IEEEbiographynophoto}

\end{document}